\documentclass[journal]{IEEEtran}

\usepackage[noadjust]{cite}

\ifCLASSINFOpdf
  \usepackage[pdftex]{graphicx}
  \graphicspath{./figures/}
  \DeclareGraphicsExtensions{.pdf,.jpeg,.png}
\else
\fi
\usepackage{amsmath}
\usepackage{algorithmic}
\usepackage{array}
\newcolumntype{$}{>{\global\let\currentrowstyle\relax}}
\newcolumntype{^}{>{\currentrowstyle}}

\ifCLASSOPTIONcompsoc
  \usepackage[font=normalsize,labelfont=sf,textfont=sf]{subfig}
\else
  \usepackage[font=footnotesize]{subfig}
\fi

\usepackage{stfloats}
\usepackage{url}

\usepackage{flushend}

\PassOptionsToPackage{hyphens}{url}
\usepackage[bookmarks=false]{hyperref}

\usepackage{verbatim}
\usepackage{fancyvrb}
\usepackage{caption}
\DeclareCaptionFormat{mylst}{#1#2#3}

\usepackage{listings}
\usepackage{courier}
\lstdefinelanguage{x86-dump}{
    sensitive=true,
    morecomment=[l]{//},
    morecomment=[s]{/*}{*/},
    keywords={push, mov, movslq, movzbl, movb, nop, retq, pop, add, test,
      setne, setl, cmpl, cmp, cmpq, repz, jge, jne, jl, js, je, jmp, shr, not,
      jbe, jns, xor, seta, neg, ja, lea, and, sub, sbb, dec, ret, imul, movl,
      cmpb, cmove}
}
\usepackage{atbegshi,ifthen,tikz}
\usetikzlibrary{fit,shapes.misc}
\tikzstyle{highlighter} = [
    yellow,
    line width = \baselineskip,
]

\newcounter{highlight}[page]

\AtBeginShipout{\AtBeginShipoutUpperLeft{\ifthenelse
        {\value{highlight} > 0}
        {\tikz[remember picture, overlay]{\foreach \stroke in
                {1,...,\arabic{highlight}} \draw[highlighter] (begin highlight
        \stroke) -- (end highlight \stroke);}}{}}}
\usepackage{booktabs}
\usepackage{tcolorbox}

\usepackage{tablefootnote}
\usepackage{threeparttable}

\usepackage{hhline}

\usepackage{dingbat}

\usepackage{float}

\usepackage[inline]{enumitem}

\newcommand{\etal}{{\it et~al}.}

\usepackage{multirow}
\newcommand{\scell}[2][c]{\begin{tabular}[#1]{@{}l@{}}#2\end{tabular}}

\begin{document}
%
\title{Input Prioritization for Testing Neural Networks}


\author{
  \IEEEauthorblockN{
    Taejoon Byun, Vaibhav Sharma, Abhishek Vijayakumar, Sanjai Rayadurgam, Darren Cofer\\
  }
  \IEEEauthorblockA{
    Department of Computer Science \& Engineering, University of Minnesota, USA\\
    Collins Aerospace, Inc. USA\\
    Email: \{taejoon, vaibhav, vijay081, rsanjai\}@umn.edu, darren.cofer@collins.com
  }
}

\maketitle


\begin{abstract}
Deep neural networks (DNNs) are increasingly being adopted for sensing and control
functions in a variety of safety and mission-critical systems such as self-driving cars, 
autonomous air vehicles, medical diagnostics and industrial robotics. Failures of such 
systems can lead to loss of life or property, which necessitates 
stringent verification and validation for providing high assurance.
Though formal verification approaches are being investigated, testing
remains the primary technique for assessing the dependability of such systems.
Due to the nature of the tasks handled by DNNs, the cost of 
obtaining test oracle data---the expected output, a.k.a. label, for a given
input---is high, which significantly impacts the amount and quality of testing
that can be performed.
Thus, prioritizing input data for testing DNNs in meaningful ways to reduce
the cost of labeling can go a long way in increasing testing efficacy. 
This paper proposes using gauges of the DNN's \emph{sentiment} derived from the
computation performed by the model, as a means to identify inputs that are likely
to reveal weaknesses.
We empirically assessed the efficacy of three such sentiment measures for
prioritization---confidence, uncertainty and surprise---and compare their
effectiveness in terms of their fault-revealing capability and retraining
effectiveness.
The results indicate that sentiment measures can effectively flag inputs
that expose unacceptable DNN behavior. For MNIST models, the average
percentage of inputs correctly flagged ranged from 88\% to 94.8\%.
\end{abstract}


%
\IEEEpeerreviewmaketitle

\section{Introduction}

Deep neural networks (DNN) are beginning to gain adoption in safety and 
mission-critical applications as a means for realizing higher levels of
autonomous operation.
However, this brings with it considerable risk because failures of such
systems can be damaging to life, property or environment. 
Recent accidents caused by autonomous vehicles or self-driving
features of automobiles~\cite{singhvi2016inside} highlight an urgent need
for rigorous yet scalable methods of assuring that DNNs will function 
in an acceptable fashion in all circumstances.
As with traditional safety-critical software, one may be able to formally
model and algorithmically verify that essential properties hold for a DNN.
However, formal verification methods are often hampered by the difficulty
of scaling up to larger models.
Thus, as is typical for all software, testing remains the most practical
and readily applicable approach for verifying DNNs.
%

%

%
%
%

It is standard practice in the DNN development process to use testing to
to evaluate the learned model. This is typically done in two phases.
The first phase is model validation with a {\it validation set}; this process is
tightly integrated with the training process to tune the hyper-parameters and
select the most appropriate model among the candidates.
The second is the testing phase, executed with a separate {\it test set},
which is independent of the training and validation sets, to evaluate the trained 
model at the final stage of development.
The development process is deemed complete only when desired properties such
as adversarial robustness or generalizability to unseen data are satisfied on
the test set.

The second phase of testing, however, can be very expensive due to the
large input space that testing has to cover, and the cost of
labeling those inputs which is needed to determine the correctness 
of the outputs  produced by a DNN over this input space.
The cost of labeling is typically much higher than that of collecting test
inputs because for many of the tasks that DNNs are designed to handle, the 
input data is abundant and easy to collect, but the oracle cannot be typically
automated--for, if otherwise, there is no need for the DNN in the first place.
Thus, it is prudent to find ways to minimize the labeling effort for new test
inputs.

One way to achieve this is to prioritize those inputs that are likely to
reveal the weakness of a trained model, so that labeling effort can be focused
only on prioritized inputs.
We hypothesize that this priority can be determined by deriving some additional
information about the computation performed by the DNN---its
\emph{sentiment}---when processing the inputs.
Higher priority inputs are those for which the DNN expresses a stronger relevant
sentiment.
In particular, we study three sentiments---\emph{confidence}, which is defined as
the predicted probability associated with the output label in DNNs that use
\emph{softmax} output layers, \emph{surprise}, which is defined as the distance
of the neuron activation pattern on an input from the activation patterns  
on the training data, and \emph{uncertainty}, which is defined for Bayesian 
Neural Networks based on the the probability distribution of the DNN's prediction.
These metrics are useful for prioritizing inputs that help us to more efficiently:
(a) address model weakness with reduced labeling cost, (b) assess model accuracy
with a reduced test suite, and (c) retrain more effectively with fewer, prioritized
inputs.
Furthermore, prioritization may also be useful in the context of run-time
monitoring of DNN components as a means to determine when to trigger system
safety or back-up mechanisms to mitigate potentially erroneous DNN outputs.

We empirically assess how these metrics perform as indicators of the test
input's value using examples of DNNs for image classification and image regression.
Our initial results show that the measures can prioritize inputs that lead
erroneous DNN outputs,
with 74.9\% to 94.8\% of average percentage of fault detection (APFD) score,
indicating that sentiment based metrics provide a meaningful basis for
prioritization. 

\section{Related Work}
Testing of DNNs is attracting consideriable interest from the research
community. For an overview, see ~\cite{braiek2018on}.
Broadly speaking, many approaches to test DNNs have focused on ways
to generate test data that expose a weakness such as lack of
robustness~\cite{pei2017deepxplore, wang2018detecting, sun2018testing, sun2018concolic}---for
example, generating an adversarial input that minimally perturbs a known input in a way
such that the expected output does not change, but leads the DNN to change its output.
Generation techniques range from introduction of adversarial
perturbations~\cite{papernot2016practical, pei2017deepxplore, sun2018concolic}
to domain-relevant transformations~\cite{tian2017deeptest, zhang2018deeproad}.
Ideas from software testing such as statement, branch~\cite{pei2017deepxplore},
condition and MC/DC coverage~\cite{sun2018testing} have been suitably modified
and adopted to define various forms of neuron 
coverage~\cite{xie2018coverage, ma2018deepguage} and show how those metrics can
be used to guide test generation~\cite{sun2018concolic, odena2018tensorfuzz, xie2018coverage}.
In these approaches, the test oracle data (expected output) is determined based,
in general, on a known metamorphic relation between the reference input and the
synthesized input.

While synthesizing test data is indeed quite useful and effective in identifying
\emph{faults}, it is, in general, not clear how to systematically identify
all these faults or determine the likelihood of the exposed faulty behavior
manifesting when the system is fielded.
Recent work in the area of adversarial input generation would seem to suggest
that DNNs provide a target-rich environment for attacks.

In the present work, we instead look for ways to rank data in some form based
on their utility for learning the DNN's weaknesses, without regard to how this
data is obtained.
This is particularly useful if we have a rich input space for which determining
the correct output is tedious.
If we aim for generalizability of the DNN---i.e. performance metrics obtained
during training and validation being truly indicative of the DNN's performance
when fielded---we need a way to determine for which inputs generalizability
\emph{may} be adversely impacted without knowing what the expected output
should be.

Works in the area of active learning studied techniques for minimizing the cost
of labeling in an interactive learning
scheme~\cite{cohn1994improving, settles2012active}, which we leverage in this
paper for test input prioritization.
Especially in a pool-based sampling~\cite{lewis1994sequential} scenario, 
training data is sampled from a large pool of unlabeled data such that
the accuracy of the model can improve most effectively with the least amount of
data.
Unlike active learning where the goal is to find training data that
provides the most information to the model under development, this work
evaluates the prioritization techniques in the context of testing where
the primary goal is to find fault-revealing inputs efficiently.

\section{White-box Test Input Prioritization}
\label{sec:prioritization}





The key idea of test input prioritization is to capture 
information available from a DNN that represents sentiments such
as confidence, uncertainty or surprise on an input presented
to the DNN.
The relative value of each test input can be judged based on the model's
sentiment, and higher priority can be assigned to uncertain or surprising inputs,
since those may more likely reveal erroneous behaviors in the model.
Although the sentiments such as uncertainty usually not provided by a
typical DNN unless explicitly modeled, multiple techniques exist that captures
the sentiments by inspecting the internal computation of the neural network.
This section introduces three such techniques.

\subsection{Softmax Output as Confidence Prediction}

Softmax is a logistic function that squashes a $K$-dimensional vector $\bf{z}$
of real values to a $K$-dimensional vector $\sigma({\bf z})$ of real values
where each entry of $\sigma({\bf z})$ is in the range $[0, 1]$ and the entries
add up to $1$: $\sigma({\bf z})_j = {e^{z_j}} / {\sum_{k=1}^{K} e^{z_k}}$, $j =
1, ..., K$.
It is typically used as the last layer of a neural network for a multi-class
classification task so that the output can represent the categorical
probability distribution of the $K$ classes.
When available, the priority score can be computed directly from the softmax
output while incurring a minimal computational overhead.
As an instantiation of the scoring function, we borrow the notion of entropy to
summarize the distribution and assign a single score to an unseen test input
${\bf x}_i$:
\begin{equation}
  {\mathit score}({\bf x}_i) = - \sum_{c=1}^{C} p_{i,c} \log p_{i,c}
\end{equation}
where $C$ is the number of output classes.
Intuitively speaking, the score is lower for a certain classification where
only one $p_{i,c}$ is high, and higher for an uncertain classification where
the predicted distribution is spread out, thus assigning high scores to inputs
that are more uncertain.

An obvious limitation of softmax-based prioritization is that it can be applied
only to classification models where a softmax layer is used.
However, a more fundamental limitation is that the predicted probability does
not reflect the model's confidence, as demonstrated by Gal and
Ghahramani~\cite{gal2016dropout} and also shown in the case of adversarial
input attacks~\cite{nguyen2014deep, goodfellow2014explaining,
subramanya2017confidence}.
For instance, an adversarially perturbed input that looks just like an ostrich
to human eyes can be classified as a panda with 99\% confidence.
These limitations call for prioritization techniques that are more reliable and
also apply to regression models.

\subsection{Bayesian Uncertainty}

Model uncertainty is the degree to which a model is uncertain about its
prediction for a given input.
An uncertain prediction can be due to a lack of training data---known as
epistemic uncertainty---or due to the inherent noise in the input data---known
as aleatoric uncertainty~\cite{kendall2017what}; but we do not distinguish the
two in this paper because they cannot be distinguished unless a neural network
is explicitly modeled to predict them as outputs~\cite{kendall2017what}.
As it is practically impossible for a machine-learning model to achieve 100\%
accuracy, model uncertainty is immensely useful for engineering a more robust
learning-enabled component.
In order to obtain model's uncertainty along with the prediction, we need
mathematically grounded techniques based on Bayesian probability theory.
We briefly introduce Bayesian neural network and a technique to approximate it using existing neural networks.
The uncertainty estimated by these techniques can then be used as scores to
prioritize test inputs.

\subsubsection{Bayesian Neural Network}

A typical (non-Bayesian) neural network has deterministic parameters that are
optimized to have fixed values.
A Bayesian neural network (BNN)~\cite{richard1991neural}, on the other hand,
treats parameters as random variables which can encode distributions.
For training, Bayesian inference~\cite{neal2012bayesian} is used to update the
posterior over the weights ${\bf W}$ given the data ${\bf X}$ and ${\bf Y}$:
$p({\bf W} | {\bf X}, {\bf Y}) = {p({\bf Y} | {\bf X}, {\bf W}) \times p({\bf W})} / {
p({\bf Y} | {\bf X}) }$, which captures the set of plausible model parameters
given the data.
To make the training of the weights tractable, the weights are often
fitted to a simple distribution such as the Gaussian, and the parameters
(mean and variance in the case of the Gaussian distribution) of the distributions
are optimized during training~\cite{graves2011practical}.
The likelihood of the prediction is often defined as a Gaussian with mean given
by the model output: $p({\bf y} | {\bf f}^{\bf W}({\bf x})) = \mathcal{N}({\bf
f}^{\bf W}({\bf x}))$ where ${\bf f}^{\bf W}({\bf x})$ denotes random
output of the BNN~\cite{kendall2017what} for an input ${\bf x}$ and $\mathcal{N}$ a normal distribution.

\subsubsection{Uncertainty in Bayesian Neural Networks}

For a classification task, the likelihood of predicting an output $c$ for an
input ${\bf x}$ is defined as:
\begin{equation} \label{eq:softmax}
  p(y = c | {\bf x}, {\bf X}, {\bf Y}) \approx \frac{1}{T} \sum_{t=1}^{T}{
  {\mathit Softmax}({\bf f}^{\bf W}({\bf x}))}
\end{equation}
with $T$ samples.
The uncertainty of the probability vector ${\bf p}$ is then summarized as the
entropy of the probability vector: $H({\bf p}) = - \sum_{c=1}^{C} p_c
\log{p_c}$.
For regression, the uncertainty is captured by the predictive variance which is
approximated as:
\begin{equation} \label{eq:var}
  {\mathit Var}({\bf y}) \approx \frac{1}{T} \sum_{t=1}^{T} {\bf f}^{\bf
  W}({\bf x})^T {\bf f}^{\bf W}({\bf x}) - E({\bf y})^T E({\bf y})
\end{equation}
with $T$ samples and the predicted mean $E({\bf y}) \approx \frac{1}{T}
\sum_{t=1}^{T} {\bf f}^{\bf W}({\bf x})$~\cite{kendall2017what}.
In other words, the predictive variance is obtained by passing the input ${\bf
x}$ $T$ times to the model ${\bf f}^{\bf W}$ and by computing the variance
among $T$ sampled outputs.



\subsubsection{Monte-Carlo Dropout as a Bayesian Approximation}

Dropout is a simple regularization technique that prevents neural network
models from over-fitting to training data~\cite{srivastava2014dropout}.
It works during the training phase by randomly dropping out some neurons in
specified layers with a given probability, so that the model parameters are
changed only for the sampled neurons.
Since the model parameters are adjusted only by an infinitesimal amount in each
iteration, the cost converges after sufficient training iterations, even with
the variance introduced by random selection of neurons.
At test time, the dropout is disabled so that every neuron participates in
making a deterministic prediction.
This simple technique is shown to be very effective in improving the
performance of neural networks on supervised learning tasks.
It was later discovered by Gal and Ghahramani~\cite{gal2016dropout} that a dropout
network can approximate a Gaussian Process~\cite{damianou2012deep}.
They proved that an arbitrary neural network with dropout applied before every
weight layer is mathematically equivalent to an approximation of a
probabilistic Gaussian process.
They also showed that any deep neural network that uses dropout layers can be
changed to produce uncertainty estimations by simply turning on the dropout at
test time (unlike the typical use case where dropout is turned off), and the
likelihood can be approximated with Monte Carlo integration.
The uncertainty of the model can then be estimated in a same way as in
Equation~\ref{eq:softmax} and~\ref{eq:var}; the only difference being that the
weight ${\bf W}$ varies by sample $t$ and follows the dropout distribution such
that $\widehat{{\bf W}}_t \sim q^{*}_{\theta}(\bf{W})$ where
$q_{\theta}(\bf{W})$ is the dropout distribution.
We refer more curious readers to the works by Gal and
Ghahraman~\cite{gal2016dropout} and Kendall and Gal~\cite{kendall2017what}.

\subsection{Input Surprise}

Surprise Adequacy (SA, in short) is a test adequacy criterion defined by
Kim~\etal~\cite{kim2018guiding} to assess the adequacy of a test suite for
testing deep learning systems.
Informally, SA is achieved when a set of test inputs demonstrates varying
degrees of model's {\it surprise}, measured by the likelihood or distance
function, relative to the training data.
The rationale is that a good test suite shall demonstrate a diverse and
representative behavior of a trained model, and that the {\it surprise} can be
a good representation of such diversity in behavior.

Unlike other coverage criteria introduced for testing neural network so
far, such as neuron coverage~\cite{pei2017deepxplore}, MC/DC-inspired
criteria~\cite{sun2018testing}, or other structural
criteria~\cite{ma2018deepguage}, SA is more fine-grained and unique in that it
can assess the quality of each input individually.
For example, SA can measure the relative surprise of an input to the training
data and give it a numeric score---the higher the score is, the more surprising
it is to the model.
%
%
Our take of SA is that a surprising input may more likely reveal an erroneous
behavior in the trained model, since a high surprise may indicate that the
model is not {\it well prepared} for the input, and thus should be given a
high score.

Kim~\etal~\cite{kim2018guiding} defined two ways of measuring the surprise.
Both ways make use of the activation trace of a neural network during a
classification. 
For the classification of every input, each neuron in the neural network gets
activated with a certain value.
The vector of activation values seen when classifying every
input can then be termed as the \textit{activation trace} of that input.
Given a set of activation traces, $A_N(T)$ for a known set of inputs $T$,
the surprise of a new input ${\bf x}$ with respect to known inputs
$T$ can be computed by comparing the activation trace of the new input
$\alpha({\bf x})$ with those from known inputs $A_N(T)$.
Kim~\etal  proposed two ways of making such comparisons.
\begin{enumerate}
\item
A probability density function can be computed for the set of activation traces
for known inputs.
For a new input, we can compute the sum of differences between its estimated
density and densities of known inputs.
The higher this sum, the more surprising the new input is.
This method is termed the \textit{Likelihood-based Surprise Adequacy~(LSA)}.
\item
Another method for comparing activation traces is to use a distance function
to create another surprise adequacy criterion called \textit{Distance-based
Surprise Adequacy~(DSA)}.
Given the set of known inputs and a new input ${\bf x}$, DSA computation 
first finds the input ${\bf x_a}$ that is the closest neighbor of ${\bf x}$
with the same predicted class as ${\bf x}$ respectively.
Next, it finds the input, ${\bf x_b}$, that is closest to $x_a$, but has a predicted
class different from the one predicted for $x$.
Next $dist_a$ and $dist_b$ is computed as:
\begin{equation} dist_a = ||\alpha_N(x) - \alpha_N(x_a)|| \end{equation}
\begin{equation} dist_b = ||\alpha_N(x_a) - \alpha_N(x_b)|| \end{equation}
with $x_a$ and $x_b$ defined as:
\begin{equation}
  x_a = \min_{\bf{D}(x_i)=c_x}||\alpha_N(x) - \alpha_N(x_i)||
\end{equation}
\begin{equation}
  x_b = \min_{\bf{D}(x_i)\in C\backslash\{c_x\}}||\alpha_N(x_a) - \alpha_N(x_i)||
\end{equation}
Finally, a value of DSA for the new input $x$ can be computed as:
\begin{equation}
DSA(x) = \frac{dist_a}{dist_b}
\end{equation}
\end{enumerate}
LSA is computationally more expensive and requires more parameter tuning than DSA.
One parameter is the small set of layers that needs to be chosen for LSA.
Activation traces for LSA will then only consist of activation values of neurons in these selected layers.
Another parameter is the value for variance used to filter out neurons whose activation values were below a certain threshold.
%
%
%
DSA, while still being sensitive to layer selection, benefits more than LSA from choosing deeper layers in the network
and has fewer parameters that need to be tuned.
For these reasons, we implemented DSA and compared it with techniques mentioned in the previous two subsections.

\section{Experiment}
\label{sec:experiment}

We experimentally assessed the efficacy of the input prioritization
techniques in two use-case scenarios.
The first use-case is labeling cost minimization.
The second use-case is retraining the model with the selected fraction of the
prioritized inputs as in active learning, which is a natural next step for
utilizing prioritized inputs.
With these scenarios in mind, we propose the following research questions.

\begin{itemize}
\item {\bf RQ1.} Can we effectively prioritize test inputs that reveal
  erroneous behavior in the model?
\item {\bf RQ2.} Can the prioritized inputs be used to retrain the model
  effectively?
\end{itemize}

The efficacy for RQ1 is measured by the cumulative percentage of error
revealing inputs after prioritization.
The efficacy for RQ2 is measured by comparing the accuracy of the model
retrained with prioritized inputs with the baseline model retrained with
randomly-selected inputs.

We answer these research questions for each prioritization technique and
compare their relative performance.
As concrete instantiation of the prioritization techniques, we compare among
the following: 1) softmax, 2) dropout Bayesian with 10 and 100 Monte-Carlo
samples, 3) Distance-based Surprise Adequacy (DSA) measured over the last one
layer and last two layers.
For the techniques that require multiple Monte-Carlo sampling, we compare
between 10 and 100 to assess the trade-off between sample size and
prioritization efficacy.
For DSA, we measure the distance of activation traces taken from the last one
layer or last two layers.
Although the efficacy of DSA can be higher when the activation traces were
taken from the middle layers of the neural networks according to
Kim~\etal~\cite{kim2018guiding}, our choice of layers were limited because of
the high space and time complexity of the DSA algorithm.
The time complexity of the DSA algorithm is quadratic to the length of the
hidden-layer output vector, and the hidden layers deeper than the last two were
typically too long to be handled efficiently given our hardware constraints.

\subsection{Systems Under Test}


To simulate a realistic testing scenario where a trained model is scrutinized
with additional test data, we chose two representative systems for image
classification and image regression.
The first system is a digit classification system trained with the 60,000
MNIST~\cite{lecun1998mnist} training dataset.
We test the system with the EMNIST~\cite{cohen2017emnist} dataset, an extension of
MNIST which is compatible to its predecessor.
The second system is called TaxiNet, which is designed for an aircraft in
ground operation to predict the distance to a center line and the heading
angle deviation from a center line while taxiing.
It is designed and developed by our industrial partner as a research prototype
to assess the applicability of learning-enabled components in the
safety-critical domain.
The data collection and training was done by ourselves.

To avoid the high cost of operating an actual aircraft in the real environment,
we collected the dataset in the X-Plane 11 simulation environment wherein the
graphics and the dynamics of the environment and the aircraft are accurately
modeled.
For a preliminary assessment, we fixed the runway to KMWH-04 and the aircraft
to be Cessna 208B Grand Caravan, while varying the position and the angle of
the aircraft together with the weather condition.
We used 40,000 samples for training with some realistic image
augmentations---such as brightness, contrast, blur, vertical affine
transformation---turned on in order to maximize the utility of the training
data and create a more robust model.

\subsection{Model Configuration}

The accuracy of a neural network depends on many factors including the amount
and quality of training data, the structure of the network, and the training
process.
As the performance of our proposed prioritization techniques may also depend on
these factors, we treated the structural configuration as an independent
variable.
However, since it is infeasible to compare the effect of all the independent
variables to the prioritization techniques, we configured a number of
representative neural networks with different structures.
We controlled the other hyper-parameters---such as learning rate and mini-batch
size---to be constant across different configurations so that the effect of the
structure alone can be studied.
The hyper-parameters are configured according to the known good practices at
the time of writing this paper so that we can objectively simulate a realistic
testing scenario.

\begin{table*}[ht]
\centering
\caption{
  Four digit classification models trained with the MNIST dataset~\cite{lecun1998mnist}
  and tested with the EMNIST dataset~\cite{cohen2017emnist}.
  %
}
\begin{tabular}{@{}crlrrr@{}}
\toprule
Model & \begin{tabular}[c]{@{}c@{}}Trainable\\ Parameters\end{tabular} & Model
Structure &
\begin{tabular}[c]{@{}c@{}}Training\\ Epochs\end{tabular} &
\begin{tabular}[c]{@{}c@{}}Validation\\ Accuracy\end{tabular} &
\begin{tabular}[c]{@{}c@{}}EMNIST\\ Test Accuracy\end{tabular} \\ \midrule
A & 594,922 & 2 Conv2D - MaxPool - 2 Conv2D - MaxPool - Flatten - Dropout - 2 Dense   &  82 & 99.16\% & 95.74\% \\
B & 177,706 & 2 Conv2D - MaxPool - Flatten - Dropout - 3 Dense  &  93 & 98.90\% & 89.66\% \\
C & 728,170 & 2 Conv2D - MaxPool - Flatten - Dropout - 3 Dense  & 138 & 98.81\% & 86.14\% \\
D & 111,514 & Dense - Dropout - 3 Dense                         & 102 & 97.74\% & 72.90\% \\
\bottomrule
\end{tabular}
\label{tbl:mnistemnist}
\end{table*}

For the digit classification task, we configured four networks as described in
Table~\ref{tbl:mnistemnist}.
For all layers except the last one, ReLu (rectified linear unit) was used as an
activation function, and L2 kernel regularization was applied to prevent the
parameters from over-fitting.
During training, we check-pointed the epoch only when the validation accuracy
(with the 10,000 validation set) improved over the previous epochs, and stopped
the training when the validation accuracy did not improve for more than twenty
consecutive epochs.

For the taxiing task, we compare two different networks named MobileNet and
SimpleNet, supplied by our industry partner.
MobileNet is a convolutional neural network inspired by
MobileNetV2~\cite{mobilenetv2}.
The structure of the network is similar to what is described in Table~2 in
Sandler~\etal's paper, and it is relatively compact in size, with 2,358,642
trainable parameters.
SimpleNet is also a convolutional neural network, with a simpler
structure, but with 4,515,338 trainable parameters.
It has five sets of convolution, batch normalization, and activation layers
back-to-back, followed by a dropout layer and four dense layers.
Both of the networks implement L2 regularization, and trained with stochastic
gradient-descent algorithm with weight decay.


\subsection{Efficacy Measure}
\label{subsec:efficacy}

\begin{figure}[h]
  \centering
  \includegraphics[width=0.7\columnwidth]{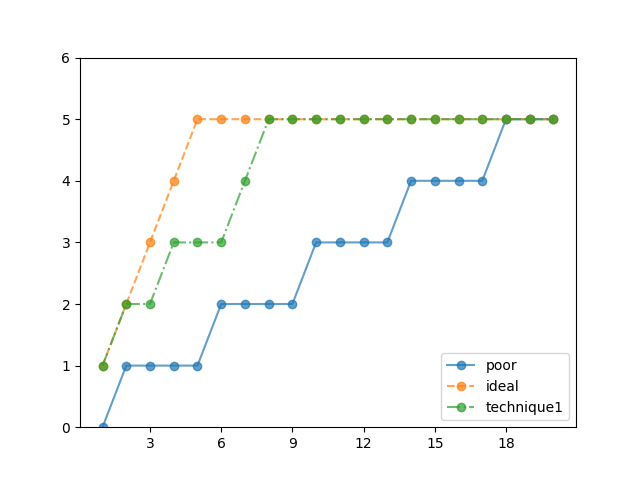}
  \caption{
    Cumulative sum of the errors found by test suites prioritized by each
    technique ({\it ideal}, {\it poor}, {\it technique1}).
    The x-axis corresponds to the test case, where the priority is higher for
    the former ones.
    The y-axis is the total number of errors found by executing the test cases
    up to $x$ test cases.
    The efficacy score of {\it technique1} is the ratio of its area under curve
    to the area under the {\it ideal} curve: $83 / 90 = 92.23$.
  }
  \label{fig:ideal}
\end{figure}

An ideal prioritization technique would consistently assign high scores to all
the error-revealing inputs and low scores to all the rest.
For example, if there were 20 prioritized test inputs among which 5 were
error-revealing, the first five inputs should all reveal errors and the rest
should not.
If we draw a graph of the cumulative sum that represents the cumulative number
of errors revealed by executing each prioritized input, the graph will be monotonically increasing until it hits 5, which is the total number of
error-revealing inputs in the given test suite (the orange line in
Figure~\ref{fig:ideal}).
A test suite without prioritization would produce a line like the blue line.
In practice, the efficacy of a prioritization technique will be somewhere
between random selection and an ideal prioritization, since it is undecidable
to predict whether the prediction is correct or not, producing a curve that
looks like the green line in Figure~\ref{fig:ideal}.
%
The efficacy of a prioritization technique can then be captured by computing
the area under curve for each technique and computing the ratio of each to the
area under the curve of the idea prioritization criterion.
This is a slight modification to Average Percentage of Fault Detected (APFD)
measure, which is typically used for measuring the efficacy of test
prioritization~\cite{rothermel1999test}.

\subsection{Implementation and Experiment Environment}

We implemented the three prioritization techniques---softmax, dropout Bayesian,
and Surprise Adequacy---in Python on top of Keras~\cite{keras}, which is one of
the most popular machine-learning libraries.
Our tool is thus compatible with any trained model that abides by Keras' Model
interface.
The surprise adequacy measurement part is implemented in C++ to better utilize
lower-level performance optimizations and thread-based parallelization.
Every feature is integrated seamlessly and provided as a Python API.
The tool is publicly available on GitHub at
\url{http://www.github.com/bntejn/keras-prioritizer}.

The experiments are performed on Ubuntu 16.04 running on an Intel i5 CPU, 32GB
DDR3 RAM, an SSD, and a single NVIDIA GTX 1080-Ti GPU.

\section{Results and Discussion}
\label{sec:result}

We ran our prioritization tool on the test datasets for all the trained models
of MNIST and TaxiNet and measured the prioritization effectiveness in terms
of misbehavior identification and retraining improvement.

\subsection{RQ1: Effectiveness of prioritization in identifying erroneous behavior}
\label{subsec:rq1}

\begin{figure*}[h]
    \centering
    \hfill
    \subfloat[MNIST-A]{
        \includegraphics[width=0.31\textwidth]{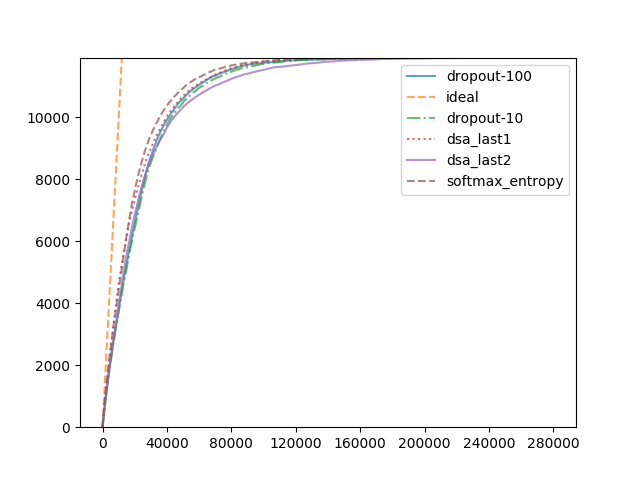}
        \label{fig:mnist_a}
    }
    \hfill
    \subfloat[MNIST-B]{
        \includegraphics[width=0.31\textwidth]{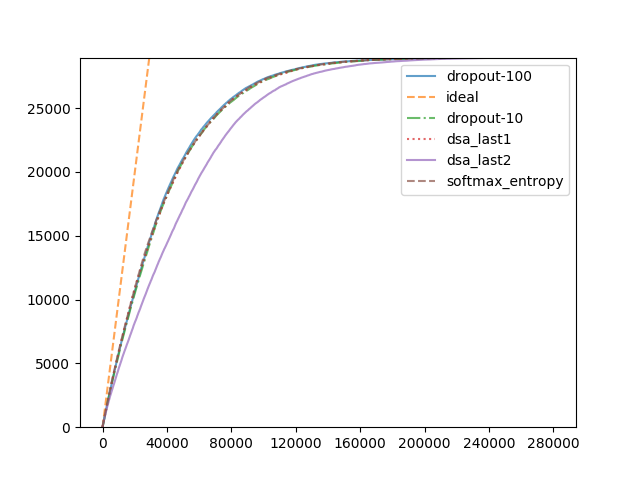}
        \label{fig:mnist_b}
    }
    \hfill
    \subfloat[MNIST-C]{
        \includegraphics[width=0.31\textwidth]{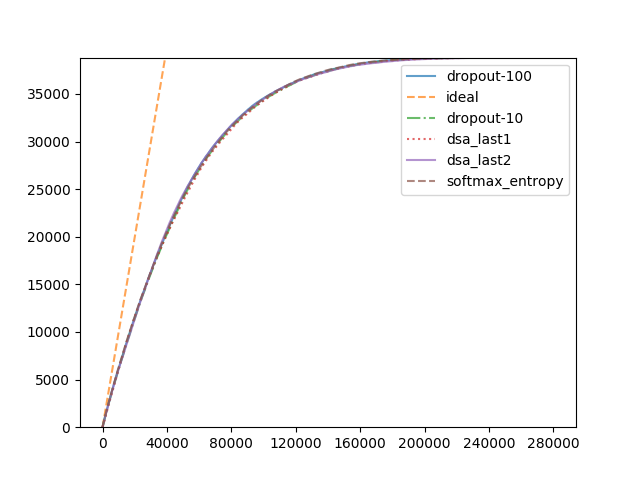}
        \label{fig:mnist_c}
    }
    \hfill
    \null  
    \hfill
    \subfloat[MNIST-D]{
        \includegraphics[width=0.31\textwidth]{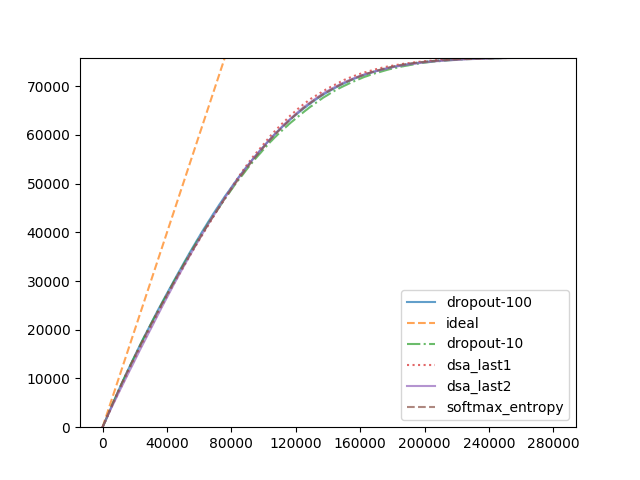}
        \label{fig:mnist_d}
    }
    \hfill
    \subfloat[TaxiNet-MobileNet]{
        \includegraphics[width=0.31\textwidth]{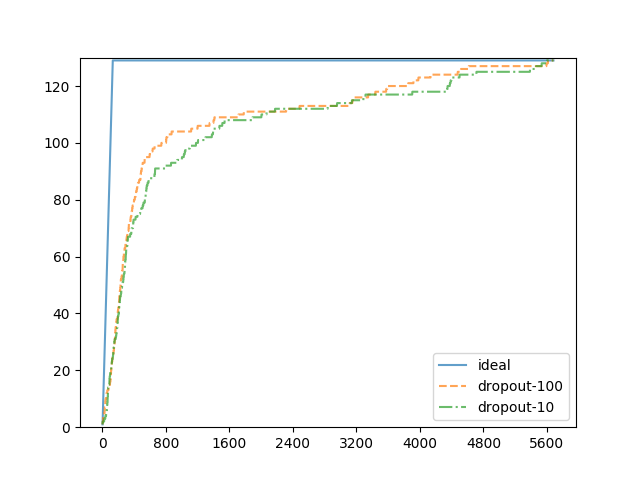}
        \label{fig:taxinet-mobile}
    }
    \hfill
    \subfloat[TaxiNet-SimpleNet]{
        \includegraphics[width=0.31\textwidth]{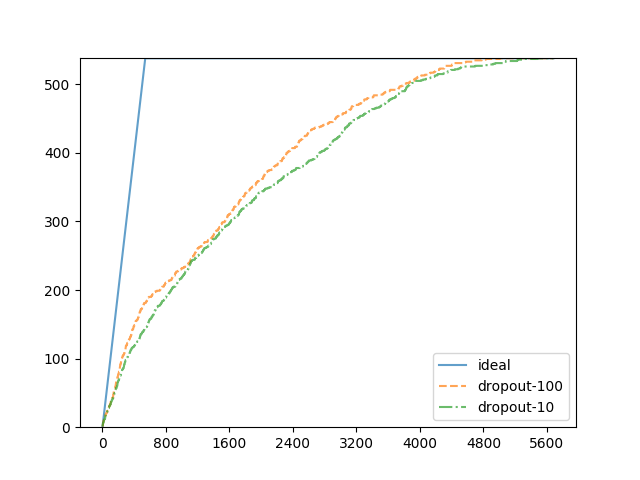}
        \label{fig:taxinet-simple}
    }
    \hfill
    \null
    \caption{
      The cumulative sum of the error revealing inputs by test inputs of
      decreasing priority:
      The x-axis represents the test cases sorted in a decreasing order of
      priority.
      The y-axis shows the cumulative sum of error-revealing inputs.
      An ideal prioritization should sort every error-revealing inputs to the
      front, drawing a highly convex curve.
      A poor prioritization, on the other hand, will produce a curve with lower
      convexity.
    }
    \label{fig:aucs}
\end{figure*}

The effectiveness of prioritization in identifying misbehavior is illustrated
in Figure~\ref{fig:aucs} and summarized in Table~\ref{tbl:auc}.
For each model we present the validation accuracy, test accuracy, and the score
of prioritization in Average Percentage of Faults Detected (APFD) as described
in Section~\ref{subsec:efficacy}.
The accuracy of classification is presented as the percentage of correct
classification, and the accuracy of regression is presented as mean absolute
error (MAE).
The MAE for TaxiNet is defined as $\frac{1}{n} \Sigma_{i=1}^n |{\bf f}^{\bf
W}({\bf x})_i - {\bf y}_i|$ where $n$---the length of the output vector---is
two for TaxiNet.
The output of TaxiNet is normalized to be between $-1$ and $1$, so the MAE is
always between $0.0$ and $1.0$ where a lower error is more desirable.
We also present the accuracy as a percentage; the correctness of an output is
determined by a fixed error threshold on MAE of $0.25$.

\begin{figure}
    \centering
    \hfill
    \subfloat[High priority input]{
        \includegraphics[width=0.23\textwidth]{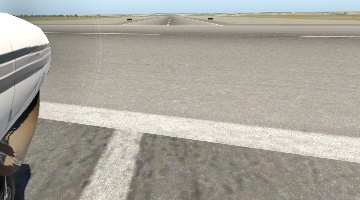}
        \label{fig:bad}
    }
    \hfill
    \subfloat[Low priority input]{
        \includegraphics[width=0.23\textwidth]{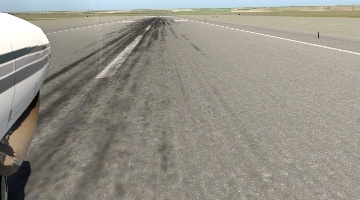}
        \label{fig:good}
    }
    \caption{
    Representative inputs of high vs. low priority: high utility score was
    assigned to inputs that produce high uncertainty---in this case, due to
    the lack of visible center line in the runway image taken around an
    intersection.
    }
    \label{fig:taxinet}
\end{figure}

\begin{table*}[t]
\caption{
  The efficacy of test input prioritization in misbehavior identification
}
\centering

\centering
\begin{tabular}{llrrrrrrr}
\toprule
\multirow{3}{*}{Dataset} & \multirow{3}{*}{Architecture}
    & \multicolumn{1}{c}{\multirow{3}{*}{Validation Accuracy}}
    & \multicolumn{1}{c}{\multirow{3}{*}{Test Accuracy}}
    & \multicolumn{5}{c}{Sentiment Measures}   \\
\cline{5-9}
 &   &   &  &
    \multirow{2}{*}{Softmax} & \multicolumn{2}{c}{Dropout}  &
    \multicolumn{2}{c}{Surprise (DSA)}  \\
    \cmidrule(lr){6-7}\cmidrule(lr){8-9}
 &   &  &  &  &
    10 & 100 &
    last1 & last2 \\
\midrule
\multirow{4}{*}{MNIST} 
  & A (CNN + Batch norm)  & 99.16\%   & 95.74\% & {\bf 94.80} & 93.20 &       93.57 &       94.26 & 92.80 \\
  & B (CNN)               & 98.90\%   & 89.66\% &       91.10 & 90.87 & {\bf 91.21} &       90.99 & 87.06 \\
  & C (CNN)               & 98.81\%   & 86.14\% &       89.30 & 89.09 & {\bf 89.35} &       88.98 & 89.26 \\
  & D (fully-connected)   & 97.74\%   & 72.90\% &       87.90 & 87.58 &      88.02  & {\bf 88.13} & 87.77 \\
\midrule
\multirow{2}{*}{TaxiNet}
  & MobileNet & 0.0394 (99.90\%) & 0.0764 (97.73\%)   & \_ & 82.84 & 86.16 & \_& \_ \\
  & SimpleNet & 0.0575 (99.66\%) & 0.1243 (90.53\%)  & \_ & 74.91 & 77.56 & \_& \_ \\
\bottomrule
\end{tabular}
\label{tbl:auc}
\end{table*}

The efficacy of prioritization presented as APFD scores ranged from $74.91$ to
$94.80$ over the different models, which suggests that test input prioritization
works in highlighting error-revealing inputs, regardless of the type of task
and the structure of the network.
Among different techniques, the efficacy of softmax, dropout Bayesian, and DSA
were all similar for the same model, but the efficacy of the dropout Bayesian
method was higher with more samples as larger samples can be used to more accurately
depict the posterior distribution.

An interesting observation is that the efficacy of the prioritization metrics
is correlated with the test accuracy of the model, or more precisely, the
difference of the validation accuracy and the test accuracy of the model.
The APFD was consistently high for the well-performing models and consistently
low for the worse-performing models, regardless of the choice of sentiment
measure.
One plausible cause for this phenomenon is covariate
shift~\cite{sugiyama2012machine}, which is a situation when the distribution
of the input data shifts from training dataset to test dataset.
A stark decrease in the test accuracy for some models suggests that the
distribution of the data shifted from the training data to test data, and some
models (such as A and B) are relatively robust to the shift while the others
are not.
Model A, for instance, implements batch normalization, a technique known to
reduce the internal covariate shift~\cite{sergey2015batch} and was more robust
to the covariate shift in input distribution, which contributed to a higher
prioritization effectiveness.
%
%
In conclusion to our first research question:
\begin{tcolorbox}
  Prioritized inputs can effectively identify erroneous behavior in a trained
  model.
  The prioritization is more effective when the model has higher test accuracy.
\end{tcolorbox}

\subsection{RQ2: Effectiveness of prioritization in retraining}
\label{subsec:rq2}

We assess the utility of the prioritized inputs when a model is retrained with
the training dataset augmented with the prioritized inputs.
This evaluation is similar to the active learning scenario but different in
that the model under test is already well-trained.
We only sample 1\% of the amount of training data from the test set, which
equals to 600 for the MNIST models and 400 for the TaxiNet models.
The baseline approach we compare against is random selection with the same
sample size---we hypothesize that prioritization techniques perform better
than the random selection.
Hyper-parameters other than the augmented training data were kept constant
in retraining runs.

\begin{table*}[ht]
\caption{
  The efficacy of input prioritization in retraining
}
\centering
\centering
\begin{tabular}{llrrrrrrr} 
\toprule
\multirow{3}{*}{Dataset} & \multirow{3}{*}{Architecture}
    & \multicolumn{1}{l}{\multirow{3}{*}{\scell{Validation Accuracy\\Baseline}}}
    & \multicolumn{1}{l}{\multirow{3}{*}{\scell{Test Accuracy\\Baseline}}}
    & \multicolumn{5}{c}{Sentiment Measures}   \\ 
\cline{5-9}
  &   & \multicolumn{1}{l}{}  & \multicolumn{1}{l}{}
    & \multicolumn{1}{c}{\multirow{2}{*}{Softmax}} 
    & \multicolumn{2}{c}{Dropout}  & \multicolumn{2}{c}{Surprise (DSA)}  \\ 
\cmidrule(lr){6-7}\cmidrule(lr){8-9}
  &   & \multicolumn{1}{l}{}  & \multicolumn{1}{l}{}  & \multicolumn{1}{c}{}
    & \multicolumn{1}{c}{10} & \multicolumn{1}{c}{100}
    & \multicolumn{1}{c}{last 1} & \multicolumn{1}{c}{last 2}  \\ 
\midrule
\multirow{4}{*}{MNIST}
  & A  & 99.23\%   & 97.48\%   & 97.95\% & 98.09 \%   & 98.06 \% & 98.18\%  & \textbf{98.44}\%  \\ 
  & B & 98.90\%   & 95.66\%   & 96.78\% & \textbf{97.41}\% & 96.26\% & 97.35\%  & 96.48\% \\
  & C  & 98.70\%   & 95.02\%   & 94.65\%  & {\bf 95.70}\% & 94.50\% & 95.02\%  & 94.45\%  \\
  & D & 97.80\%   & 90.57\%   & 89.93\% & 87.58\% & 88.26\% & 91.87\% & \textbf{92.00}\%  \\
\midrule
\multirow{2}{*}{TaxiNet}
    & MobileNet & 0.0336 (100.00\%) & 0.0364 (99.86\%) & \_  & 99.84\%  & {\bf 99.89}\% & \_& \_ \\
    & Simple    & 0.0502 (99.85\%) & 0.0522 (99.05\%) & \_  & {\bf 99.71}\% & 99.56\% & \_& \_ \\
\bottomrule
\end{tabular}
\label{tbl:retrain}
\end{table*}

Table~\ref{tbl:retrain} shows that the relative efficacy of retraining follows
a similar trend to the error-revealing efficacy presented in Table~\ref{tbl:auc}.
The prioritized inputs could improve the accuracy of the retrained models more
effectively than randomly sampled inputs in most cases, and the efficacy was
more pronounced for MNIST model A and B than in model C and D.
When model B and C are compared, model B consistently performed better when
retrained with prioritized inputs while model C almost always performed worse
when retrained with with randomly sampled inputs.
One hypothetical explanation could be that a well-architectured DNN model with
a better generalization benefits more from learning the corner-cases, whereas
an unoptimal DNN learns more from general cases.
However, the exact reason for this phenomenon cannot be drawn from the limited
experiment---a future investigation is necessary.
In conclusion to the second research question:
\begin{tcolorbox}
  Sentiment measures can prioritize inputs that can augment the training
  dataset with which a better accuracy can be achieved.
  But random sampling was found more effective for the models that achieve
  low test accuracy.
\end{tcolorbox}

\subsection{Threats to Validity}

In the experiment, we evaluated the sentiment measures with both an image
classification task and an image regression task, and configured several
DNNs with various structural features.
Despite our effort, the representativeness of the configured DNNs were
inevitably limited in number and variety, and our empirical findings
might not generalize to other types of DNNs such as recurrent neural
nets.
Nevertheless, the DNNs we evaluated are of realistic sizes and implement
some of the most widely used techniques that are applied in practical deep
learning practices~\cite{goodfellow2016deep}.

The second experiment for answering RQ2 was performed without the
statistical rigor required for hypothesis testing due to the prohibitive
cost of retraining a large model multiple times.
We present the result and the finding as a preliminary assessment of the
sentiment measures in the context of testing which calls for more rigorous
empirical assessment in future work.

\section{Conclusion and Future Work}

This paper presented techniques for mitigating the oracle problem in testing
DNNs by prioritizing error-revealing inputs based on white-box measures of
DNN's sentiment---softmax confidence, Bayesian uncertainty, and input surprise.
We evaluated the three techniques on two example systems for image classification
and image regression, and multiple versions of the DNNs configured with different
architectures.
The experiment showed that the sentiment measures can prioritize error-revealing
inputs with an average fault-detection rate of 74.9 to 94.8, indicating that
input prioritization based on sentiment measures is a viable approach for
effectively identifying weakness of trained models with reduced labeling cost.

We firmly believe that more attention should be paid to techniques that can facilitate
field testing of safety-critical DNNs, which can be a laborious process and
that test prioritization is an important step towards that goal, providing practical
utility and good scalability.
Further research is still warranted for assessing the representativeness and 
completeness of test sets with respect to the operational environments of 
DNN-based systems.



\section*{Acknowledgment}

This work is supported by AFRL and DARPA under contract FA8750-18-C-0099.


\bibliographystyle{IEEEtran}
\bibliography{uncertainty}
%

\end{document}